\documentclass[twocolumn,showpacs,preprintnumbers,amsmath,amssymb,aps,superscriptaddress,prb,10pt]{revtex4-2}
\usepackage{textcomp}
\usepackage{graphicx}
\usepackage{subfigure}
\usepackage{bm}
\usepackage[final=true]{hyperref}

\newcommand{\m}{\mathbf}

\newcommand{\ve}{\varepsilon}
\newcommand{\be}{\begin{eqnarray}}
\newcommand{\ee}{\end{eqnarray}}
\newcommand{\nn}{\nonumber}

\begin{document}

\title{Easy-plane antiferromagnet in tilted field: gap in magnon spectrum and susceptibility}

\author{Artemiy S. Sherbakov}
\email{nanoscienceisart@gmail.com}
\affiliation{National Research Center ``Kurchatov Institute'' B.P.\ Konstantinov Petersburg Nuclear Physics Institute, Gatchina 188300, Russia}

\author{Oleg I. Utesov}
\email{utiosov@gmail.com}

\affiliation{Saint Petersburg State University, Ulyanovskaya 1, St. Petersburg 198504, Russia}
\affiliation{National Research Center ``Kurchatov Institute'' B.P.\ Konstantinov Petersburg Nuclear Physics Institute, Gatchina 188300, Russia}
\affiliation{St. Petersburg School of Physics, Mathematics, and Computer Science, HSE University, St. Petersburg 190008, Russia}

\begin{abstract}

Motivated by recent experimental data on dichloro-tetrakis thiourea-nickel (DTN) [Soldatov \emph{et al}, Phys. Rev. B {\bf 101}, 104410 (2020)], a model of antiferromagnet on a tetragonal lattice with single-ion easy-plane anisotropy in the tilted external magnetic field is considered. Using the smallness of the in-plane field component, we analytically address field dependence of the energy gap in ``acoustic'' magnon mode and transverse uniform magnetic susceptibility in the ordered phase. It is shown that the former is non-monotonic due to quantum fluctuations, which was indeed observed experimentally. The latter  is essentially dependent on the ``optical'' magnon rate of decay on two magnons. At magnetic fields close to the one which corresponds to the center of the ordered phase, it leads to experimentally observed dynamical diamagnetism phenomenon.

\end{abstract}

\maketitle
\section{Introduction}

Quantum phase transitions and quantum phases of matter are among the most extensively studied subjects of contemporary condensed matter physics~\cite{sachdev2011}. In this context, quantum magnets play an important role in revealing versatile physics~\cite{Mila, giamarchi2008bose, zheludev2013dirty}. Importantly, in such compounds one can add impurities in more or less controllable way~\cite{oosawa2002random,yu2012bose,huvonen2012}, which allows studying effects of quenched disorder and so-called ``dirty-boson'' systems~\cite{Fisher,theorem}. Moreover, such elusive phases as Bose and Mott glasses can be observed in disordered quantum magnets\cite{yu2012bose}.

\begin{figure}[h]
  \centering
  \includegraphics[width=6cm]{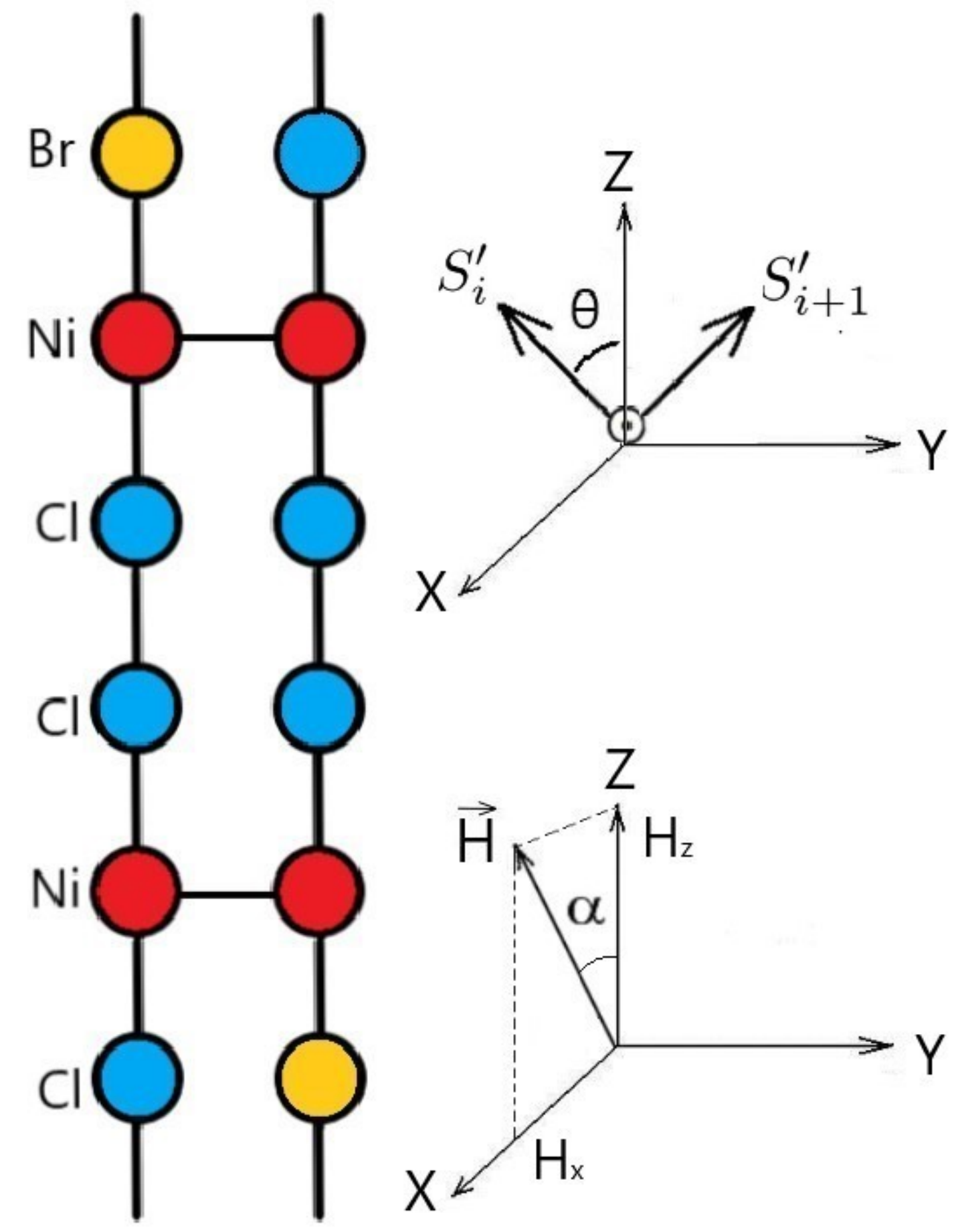}
  \caption{In the present paper, a model of an antiferromagnet relevant to DTN and DTNX is considered. It consists of relatively weakly coupled antiferromagnetic chains. There is also large easy-plane single-ion anisotropy. In moderate external fields along the $z$ axis, the spins are canted towards the field direction. Moreover, the slightly tilted towards the $x$ axis field leads to rotational symmetry breaking and to an average $S^x$ component, which results in the emergence of the gap in the magnon spectrum.
\label{fig:localcor}}
\end{figure}

Particular compounds DTN [dichloro-tetrakis thiourea-nickel, NiCl$_2$-4SC(NH$_2$)$_2$] and its bromine-doped counterpart DTNX have been studied for last several decades (see, e.g., Refs.~\cite{paduan2004,Zvyagin2007,sizanov2011antiferromagnet,Sizanov2011,yu2012bose,Povarov2017, Orlova_2018} ). However, they are still able to reveal peculiar features. Their properties can be described using the model of weakly coupled antiferromagnetic (AF) chains with spin \mbox{$S=1$} and strong single-ion easy-plane anisotropy~\cite{zapf2006}, see Fig.~\ref{fig:localcor}. The latter is responsible for DTN interesting magnetic properties. In the external magnetic field $h$ along the tetragonal axis, at $h<h_{c1}$ the system is in singlet paramagnetic phase, at $h_{c1}<h<h_{c2}$ ordered canted AF (CAF) phase emerges, whereas at $h>h_{c2}$ the fully polarized gapped state can be observed~\cite{paduan2004,zapf2006}. Critical fields $h_{c1}, \, h_{c2}$ and their vicinity can be discussed in terms of elementary excitations (triplons and magnons, respectively) Bose-Einstein condensation~\cite{batyev1984,batyev1985,zapf2006,yin2008,giamarchi2008bose,sizanov2011antiferromagnet}. The intermediate CAF phase not very close to the critical fields can be described as a usual magnetically ordered phase and studied, e.g., by means of electron spin resonance technique~\cite{Zvyagin2007,smirnov,smirnov2}. Interestingly, such experiments reveal, for instance, nontrivial field variation of the ``optical'' (high-energy) magnon frequency.  In our previous work~\cite{sherutesov} it was explained in the framework of conventional diagrammatic $1/S$ expansion, where the role of strong quantum fluctuations was highlighted.

In the present paper, we continue our theoretical discussion~\cite{sherutesov}. Here we address the case of the tilted magnetic field realized experimentally~\cite{smirnov2}. Using the Kubo formalism for linear response and $1/S$ expansion, we show that induced by the tilted field in-plane spin component inherits peculiar field-behavior of the optical magnon frequency. This leads to experimentally observed field dependence of the gap in the ``acoustic'' (low-energy) branch of the spectrum. Moreover, we show that the optical branch near the center of the ordered phase acquires significant damping due to two-magnon processes. So, our theory quantitatively supports qualitative discussion of the dynamic diamagnetism phenomenon observed in Ref.~\cite{smirnov2}.

The rest of the paper is organized as follows. In Sec.~\ref{SGap} we formulate the model, briefly recall our previous results, and employ our approach for the gap in acoustic branch derivation. Section~\ref{SMag} is devoted to homogeneous transverse susceptibility in the tilted field and the dynamical diamagnetism effect. We present our conclusions in Sec.~\ref{SConc}. Three appendices contain cumbersome formulas and some other details of calculations.

\section{Gap in magnon spectrum}
\label{SGap}

\subsection{Problem formulation}

We consider a model of an antiferromagnet with single-ion easy-plane anisotropy in the canted external magnetic field on a simple tetragonal lattice\cite{Lopez1963}, see Fig.~\ref{fig:localcor}. For definiteness, we take the magnetic field in the $xz$ plane ($xy$ is the easy plane). The corresponding Hamiltonian reads
\begin{equation}\label{hamiltonian}
  \mathcal{H} = \dfrac{1}{2} \sum_{<i,j>} J_{ij}\textbf{S}_{i} \cdot \textbf{S}_{j}+D\sum_{i}({S_{i}^{z}})^{2}-h_{z}\sum_{i}S_{i}^{z}-h_{x}\sum_{i}S_{i}^{x}.
\end{equation}
Here the external magnetic field is taken in the energy units ($ h = g \mu_B |\m{H}|$) and $D>0$ indicates the easy-plane anisotropy. Since our aim is to address experimentally relevant case~\cite{smirnov,smirnov2} of small tilt angles $\alpha$, we will treat the part of the Hamiltonian with $h_x \approx \alpha h$ as a small perturbation. Hence, the starting point of our analysis will be a theory from our previous paper~\cite{sherutesov}. Let us briefly recall its main findings.

We considered the case of the external field along the tetragonal axis. We mainly discussed the ``optical'' magnon branch and its energy in the center of the Brillouin zone $\Delta$ (which can be probed by ESR~\cite{smirnov,smirnov2}) employing Holstein-Primakoff representation of spin operators via bosonic ones~\cite{holstein} in the alternating coordinate frame suitable for the canted antiferromagnetic ordering~\cite{Hamer2010}. Using quasiclassical $1/S$-series diagrammatic approach, we found a strong renormalization of the magnon spectrum due to the first order in $1/S$ quantum corrections, which lead to experimentally observed~\cite{smirnov,smirnov2} nonmonotonous dependence $\Delta(h)$.





Despite being small, the transverse magnetic field $h_{x}$ has an important effect on the system: it breaks the rotational symmetry of the easy $xy$ plane. This results in the disappearance of the Goldstone mode, the corresponding magnon branch becomes gapped and will be called ``quasi-Goldstone'' below.

Our main goal is to analytically describe the properties of the emergent gap (as a function of $h$ and $\alpha$) and the magnetic susceptibility of the system. For this sake, we start from the ground state at $\alpha=0$ and immediately observe that $h_x$ terms in the Hamiltonian~\eqref{hamiltonian} provide linear in bosonic operators terms (equivalent to the ground state instability), which in the reciprocal space read
\be \label{hamlin}
  \frac{h_{x}\sqrt{SN}}{\sqrt{2}}\left(a^{\dagger}_{\mathbf{k}=0}+a_{\mathbf{k}=0}\right),
\ee
see Appendix~\ref{appendhx0} for some details. We eliminate these terms using the following shift in operators:
\begin{equation} \label{intrho}
  a_{\mathbf{k}=0}=b_{\mathbf{k}=0}+\rho,\quad a^{\dagger}_{\mathbf{k}=0}=b^{\dagger}_{\mathbf{k}=0}+\rho.
\end{equation}
Parameter $\rho$ is related to the linear response of the system to $h_x$ and can be found directly from the Hamiltonian or using the Kubo susceptibility. In any case, we found that $\rho \propto \alpha$. Importantly, $\rho \neq 0$ leads to new terms in the Hamiltonian, which will be used for discussing the gap in the quasi-Goldstone mode and the magnetic susceptibility in the tilted field. Note that the smallness of $\alpha$ allows us to omit terms $\propto \rho^n, n>2$ when recalculating the Hamiltonian for nonzero $h_x$.


\subsection{Calculation of $\rho$}

Here we derive an analytical expression for the crucial parameter $\rho$. Importantly, it is simply related to the average $S^x$ value, which is the same for all spins. Explicitly:
\begin{multline}\label{sxrho}
  \langle S^{x}\rangle=\frac{1}{N}\sum_{i}\langle S^{x}_{i}\rangle=\frac{\sqrt{S}}{\sqrt{2}N}\sum_{i}\langle a^{\dagger}_{i}+a_{i}\rangle \\
  =\frac{\sqrt{S}}{\sqrt{2N}}\langle a^{\dagger}_{\m{k=0}}+a_{\m{k=0}}\rangle=\sqrt{2}S\frac{\rho}{\sqrt{SN}}.
\end{multline}
Parameter $\rho$ can be found by making linear terms in the Hamiltonian to be zero [see Eq.~\eqref{hamlin}] using the shift~\eqref{intrho}. However, in order to take into account quantum corrections, we will use a more general method of the Kubo susceptibility:
\begin{eqnarray}\label{kubo}
&\chi=\langle S^{x}(\m{q},\omega)S^{x}(-\m{q},\omega)\rangle \Bigr|_{\m{q}=0,\omega=0,h_{x}=0}=\nonumber\\
&=\frac{S}{2}\langle (a^{\dagger}_{\m{q}}+a_{-\m{q}})( a^{\dagger}_{-\m{q}}+a_{\m{q}})\rangle_{\omega}\Bigr|_{\m{q}=0,\omega=0,h_{x}=0}=\nonumber\\
&=-\frac{S}{2}(F_{k=0}+G_{k=0}+F^{\dagger}_{k=0}+G_{k=0}).\label{chi}
\end{eqnarray}
So, $\chi$ is the sum of retarded magnon Green's functions taken at $k = (\omega=0, \mathbf{k}=0)$, see Appendix~\ref{appendhx0} for details. The latter can be obtained at $h_x=0$ using the conventional diagrammatic approach.

In the linear spin-wave theory we have
\be
  \chi_{0}=\frac{1}{2SJ_{\m{0}}}, \label{chil}
\ee
which yields
\be
  \langle S^{x}\rangle=\dfrac{h_{x}}{2SJ_{\m{0}}}, \label{sxl} \\
  \frac{\rho}{\sqrt{NS}}=\dfrac{h_{x}}{2\sqrt{2}SJ_{\m{0}}}. \label{linrho}
\ee
One sees that the linear spin-wave theory predicts constant susceptibility. 

\begin{figure}[h]
  \centering
  \includegraphics[width=8cm]{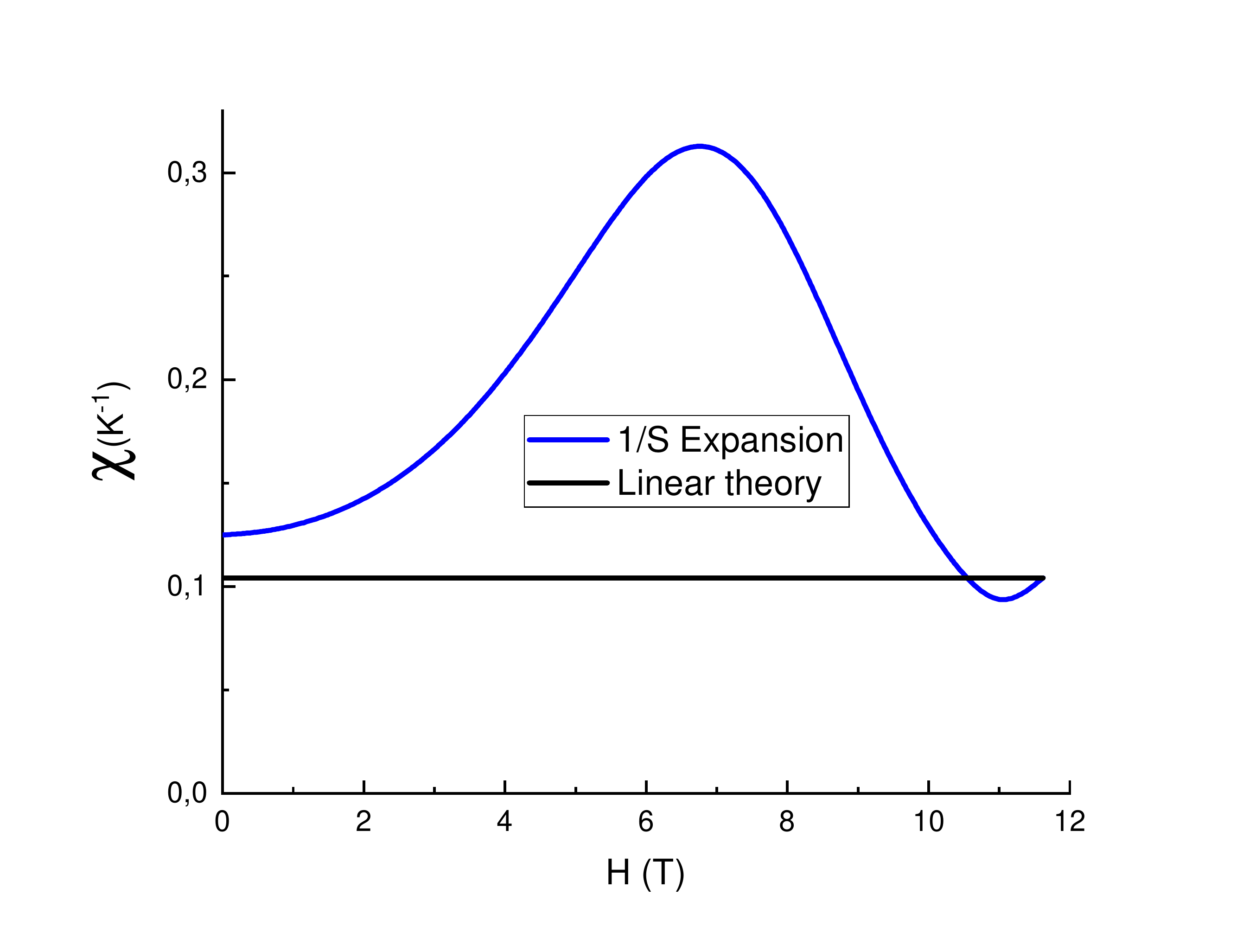}
  \caption{Comparison between magnetic susceptibility calculated in the linear theory $\chi_0$ (black line) and the one, including $1/S$ corrections to Green's functions, $\chi_{1/S}$, (blue line) as functions of magnetic field $h$. Parameters $\alpha=\pi/60$, $D=8$ K, $J_c=2$ K,  $J_{a}=0.2$ K were used.
\label{fig:chilq}}
\end{figure}

\begin{figure}[h]
  \centering
  \includegraphics[width=8cm]{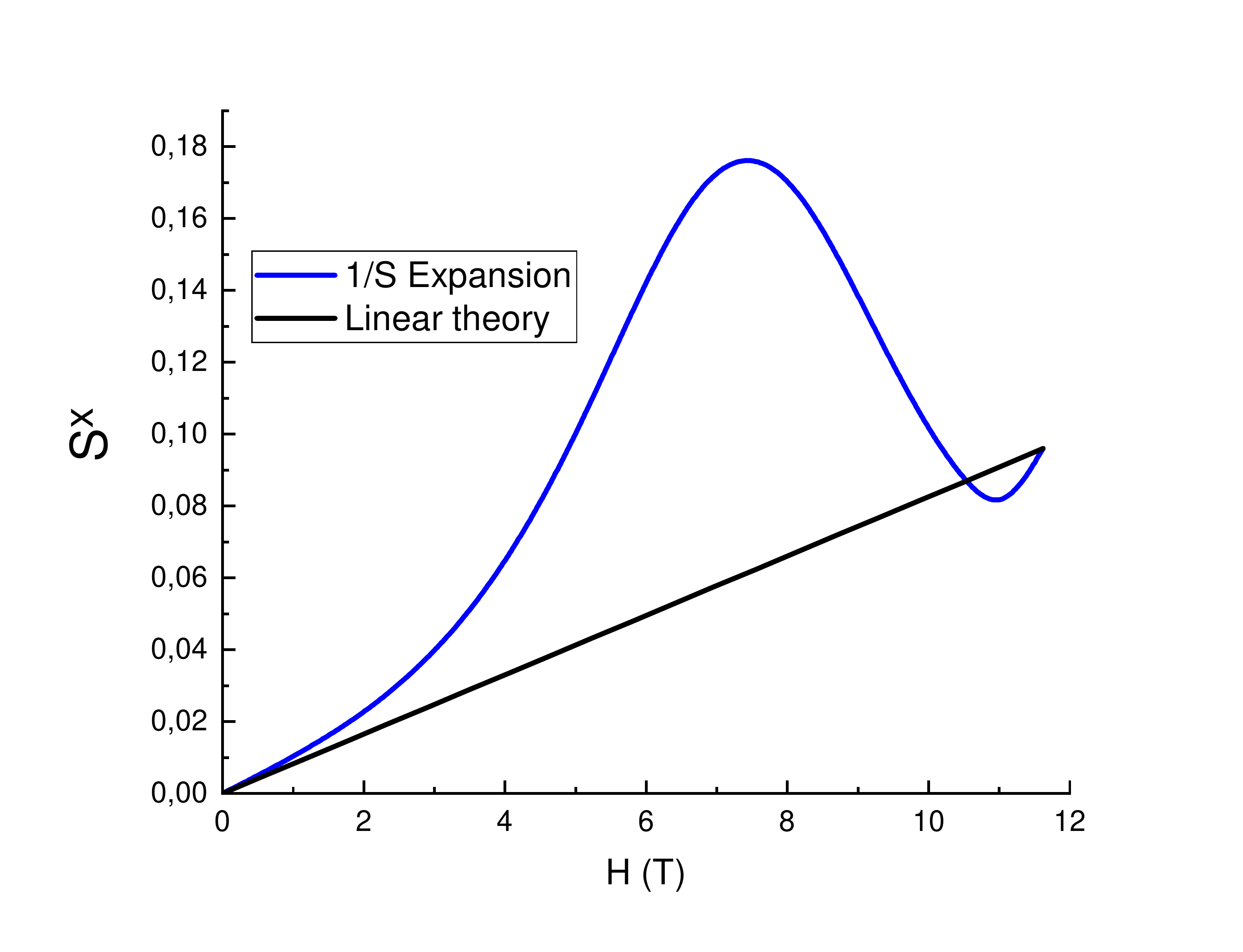}
  \caption{The dependence of the average spin component $S^{x}$ on $h$ in the tilted field case, calculated using Kubo formalism in linear theory (black line) and including quantum corrections (blue line). Parameters $\alpha=\pi/60$, $D=8$ K, $J_c=2$ K, $J_{a}=0.2$ K were used.
\label{fig:sx}}
\end{figure}

When $1/S$ corrections are taken into account, the susceptibility behavior becomes much more interesting. Using Eq.~\eqref{chi} with renormalized by magnon interaction Green's functions, we obtain
\be
  \chi_{1/S}(h)=\frac{P(h)}{2\Delta^2(h)}, \label{chis}
\ee
where 
\begin{eqnarray}
   P(h)&=&2E_{\m{k}}-2B_{\m{k}}+\Sigma_{k}+\Sigma_{-k}-2\Pi_{k} \Bigr|_{k=0}\label{Ph},\\
   \Delta(h) &=& \varepsilon_{\mathbf{k}}+(\Sigma_{k}-\Sigma_{-k})/2 \nonumber\\
  && + \left[E_{\m{k}}(\Sigma_{k}+\Sigma_{-k})-2B_{\m{k}}\Pi_{k}\right]/2\varepsilon_{\mathbf{k}} \Bigr|_{k=0}.
\end{eqnarray}
Here $\ve_{\m{k}}$ is the magnon spectrum of the linear theory and, as one can see, we use the on-shell approximation for self-energy parts. Importantly, $\Delta(h)$ is renormalized energy of the optical magnon with $\m{k}=0$, which was studied in detail in Ref.~\cite{sherutesov} and has nonmonotonic behavior as a function of $h$ due to the $1/S$ corrections. This effect is also pronounced in the susceptibility and $\langle S^x \rangle$, as it is shown in Figs.~\ref{fig:chilq} and~\ref{fig:sx} where the linear spin-wave theory results are contrasted to the ones, obtained using the first $1/S$ corrections. We report drastic differences in the corresponding results, which are inherited from $\Delta(h)$.

\subsection{Gap at $h_x \neq 0$.}

After the shift in Bose-operators~\eqref{intrho}, multiple terms \mbox{$\propto \rho^n, \, n\geq 1$} emerge in the Hamiltonian in addition to those discussed in Appendix~\ref{appendhx0}. Some details are presented in Appendix~\ref{appendhx}.

Linear in Bose-operators terms are canceled by proper choice of $\rho$ (see previous Subsec.). We also neglect ``decay'' terms $\delta\mathcal{H}_3$ (and the higher order in Bose-operators ones) because they provide small in $1/S$ corrections. Moreover, the subject of the calculations below is magnon with $\m{k}=\m{k}_0$ and of minimal energy, which cannot acquire a finite lifetime due to such processes.

So, we focus on corrections to the bilinear part of the Hamiltonian. Importantly, they include normal and umklapp parts, which read
\be \label{Hnorm}
  \delta\mathcal{H}^{N}_{2}&=& S\sum_{\mathbf{k}} \overline{E}_{\mathbf{k}}b_{\mathbf{k}}^{\dagger}b_{\mathbf{k}}  + S \sum_{\mathbf{k}} \overline{B}_{\mathbf{k}} \frac{b_{\mathbf{k}}b_{-\mathbf{k}}+b_{\mathbf{k}}^{\dagger}b_{-\mathbf{k}}^{\dagger}}{2}, \\ \label{Humk}
  \delta\mathcal{H}_{2}^{U}&=& iS\sum_{\mathbf{k}} X_{\mathbf{k}}\left(b_{\mathbf{k}}^{\dagger}b_{\mathbf{k}-\mathbf{k_{0}}}-b_{\mathbf{k}-\mathbf{k_{0}}}^{\dagger}b_{\mathbf{k}} \right)  \\
\nonumber&&+ i S \sum_{\mathbf{k}}  Y_{\mathbf{k}} \frac{b_{\mathbf{k}}b_{\mathbf{k_{0}}-\mathbf{k}}-b_{\mathbf{k}}^{\dagger}b_{\mathbf{k_{0}}-\mathbf{k}}^{\dagger}}{2},
\ee
respectively. Explicit formulas for the coefficients are presented in Appendix~\ref{appendhx}.

\begin{figure}
  \centering
  \includegraphics[width=8cm]{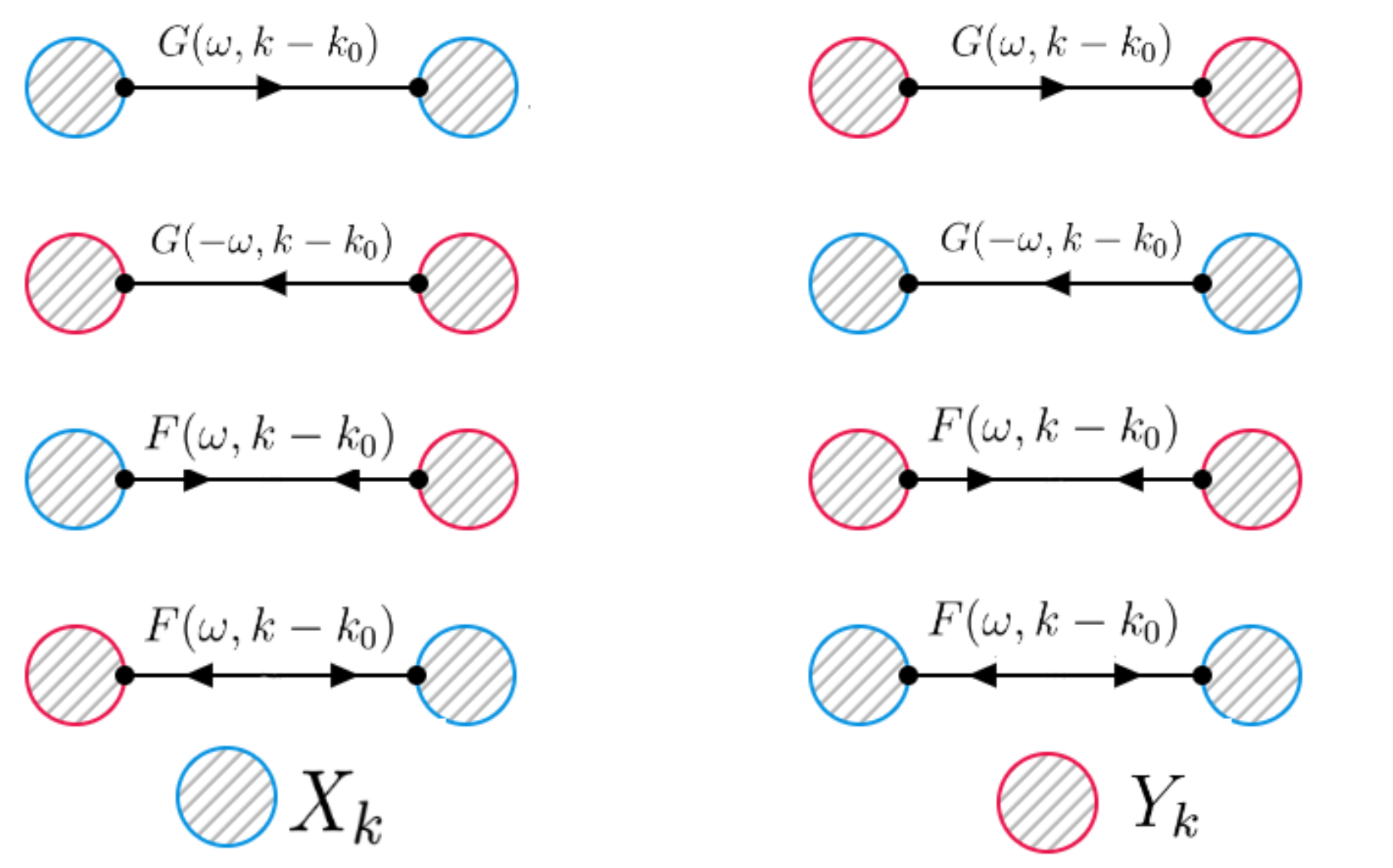}
  \caption{Umklapp terms-induced corrections to the self-energy parts $\Sigma^{U}$ (left) and $\Pi^{U}$ (right). $X_\m{k}$ and $Y_\m{k}$ denote the corresponding umklapp amplitudes, see Eqs.~\eqref{a} and~\eqref{y}.
\label{fig:Umklapp}}
\end{figure}

Next, one can see that the normal corrections can be directly plugged into the equation for the magnon spectrum of the linear theory~\eqref{speclin}, whereas the umklapps require some additional calculations. They contribute to both normal and anomalous self-energy parts; corresponding diagrams are shown in Fig.~\ref{fig:Umklapp}.

However, at $\mathbf{k=k_{0}}$ only the normal terms contribute to the spectrum since the umklapp corrections cancel each other (see Appendix~\ref{appendhx}). So, the equation for the gap in the spectrum reads:
\begin{eqnarray}\label{deltahx}
\omega_{\mathbf{k=k_{0}}} =\Delta_{h_{x}}=\sqrt{2E_{\m{k}_{0}}(\overline{E}_{\m{k}_{0}}+\overline{B}_{\m{k}_{0}})}
\end{eqnarray}
Using the expressions for $\overline{E}_{\m{k}_{0}}$~\eqref{z} and $\overline{B}_{\m{k}_{0}}$~\eqref{m} and taking the value of parameter $\rho$ calculated within the Kubo formalism, we get
\begin{eqnarray}
\Delta_{h_{x}}=\sqrt{2E_{\m{k_{0}}}A_{h}}\dfrac{S\rho}{\sqrt{NS}},
\ee
where
\be
A_{h}=\dfrac{h_{x}\sqrt{NS}}{S\rho}+8U_{\m{0k_{0}0-k_{0}}}+12V_{\m{0k_{0}0}}+8U_{\m{k_{0}k_{0}00}}+12V_{\m{0k_{0}k_{0}}}\label{Ah}.\nn \\
 \end{eqnarray}
The first term in Eq.~\eqref{Ah} is much smaller than the others and it is significant only near the critical fields $h_{c1}$ and $h_{c2}$. In the intermediate field region (on which we mainly focus) it can be safely neglected. Finally, we obtain the following result for the gap value:

 \begin{eqnarray}
 \Delta_{h_{x}}=\sqrt{E_{\m{k_{0}}}A_{h}}\langle S^{x}\rangle. \label{gapspin}
 \end{eqnarray}
 
\begin{figure}
  \centering
  \includegraphics[width=8cm]{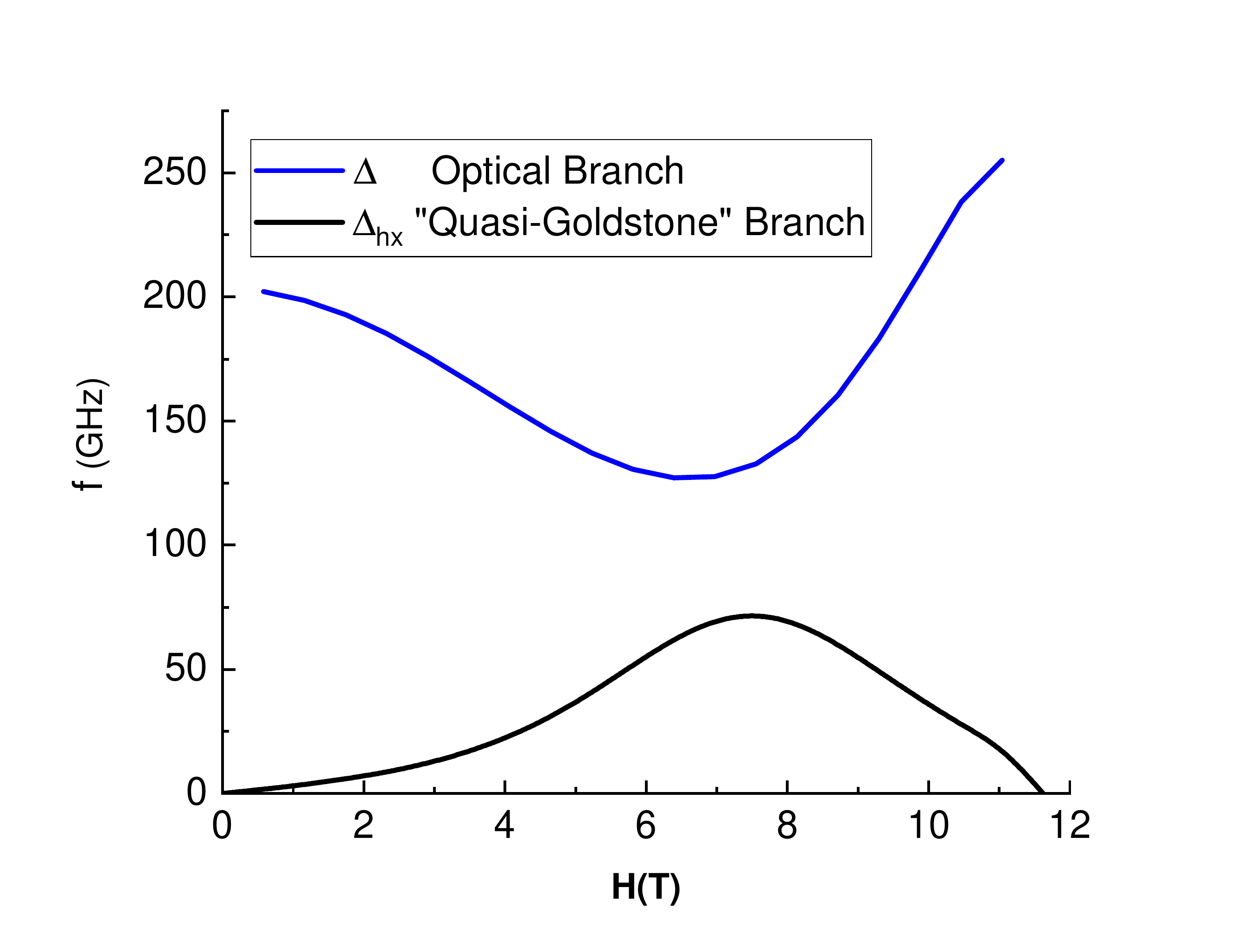}
  \caption{Field dependencies of frequencies of ``quasi-Goldstone'' ($\m{k}=\m{k}_0$; black curve) and ``optical'' ($\m{k}=\m{0}$; blue curve) magnons. Parameters $\alpha=\pi/60$, $D=8.0$K, $J_c=2$K, and $J_{a}=0.2$K were used.
\label{fig:branches}}
\end{figure}

Results of our calculations are presented in Fig.~\ref{fig:branches} for a particular parameter set. Note that the gap is proportional to the average $S^x$ value (cf. Fig.~\ref{fig:sx}). Moreover, its behavior is similar to the one, observed experimentally~\cite{smirnov,smirnov2}.



\section{Magnetic susceptibility}
\label{SMag}

Here we discuss the uniform transverse magnetic susceptibility of the system in the tilted external field. A crucial point here is the significant damping of the optical magnon with $\m{k}=0$, which is addressed below.

\subsection{Magnon damping}

In our previous work~\cite{sherutesov}, we were particularly interested in the energy of the optical magnon with $\m{k}=0$. For our current purposes, it's damping also becomes important. Theoretically, the finite magnon lifetime is connected with the possibility for it to decay; in our case on two other magnons. 

Explicitly, the damping comes from the imaginary part of magnon Green's function self-energies [see Eq.~\eqref{Green}]. Moreover, here we can neglect small $h_x$-induced terms in the Hamiltonian. Separating the imaginary part of the corresponding denominator from the real one, we get it in the first order in $1/S$:
\begin{eqnarray}\label{Gamma}
\Gamma(k) = \mathrm{Im} \left[\dfrac{\Sigma_{k}-\Sigma_{-k}}{2}+\dfrac{E_\m{k}(\Sigma_{k}+\Sigma_{-k})-2B_\m{k}\Pi_{k}}{2\varepsilon_{k}}\right].\nonumber\\
\end{eqnarray}
Corresponding imaginary parts of normal and anomalous self-energy parts are due to $\mathcal{H}_3$ terms [see Eq.~\eqref{h3}] where a magnon with momentum $\mathbf{k}$ can decay on magnons with momenta $\mathbf{k-p+k_{0}}$ and $\mathbf{p}$. Below, we only discuss the damping of magnon with $\m{k}=\m{0}$ and use the on-shell approximation. The corresponding damping is $h$-dependent, we denote it as $\Gamma(h)$.


\begin{figure}[h]
  \centering
  \includegraphics[width=8cm]{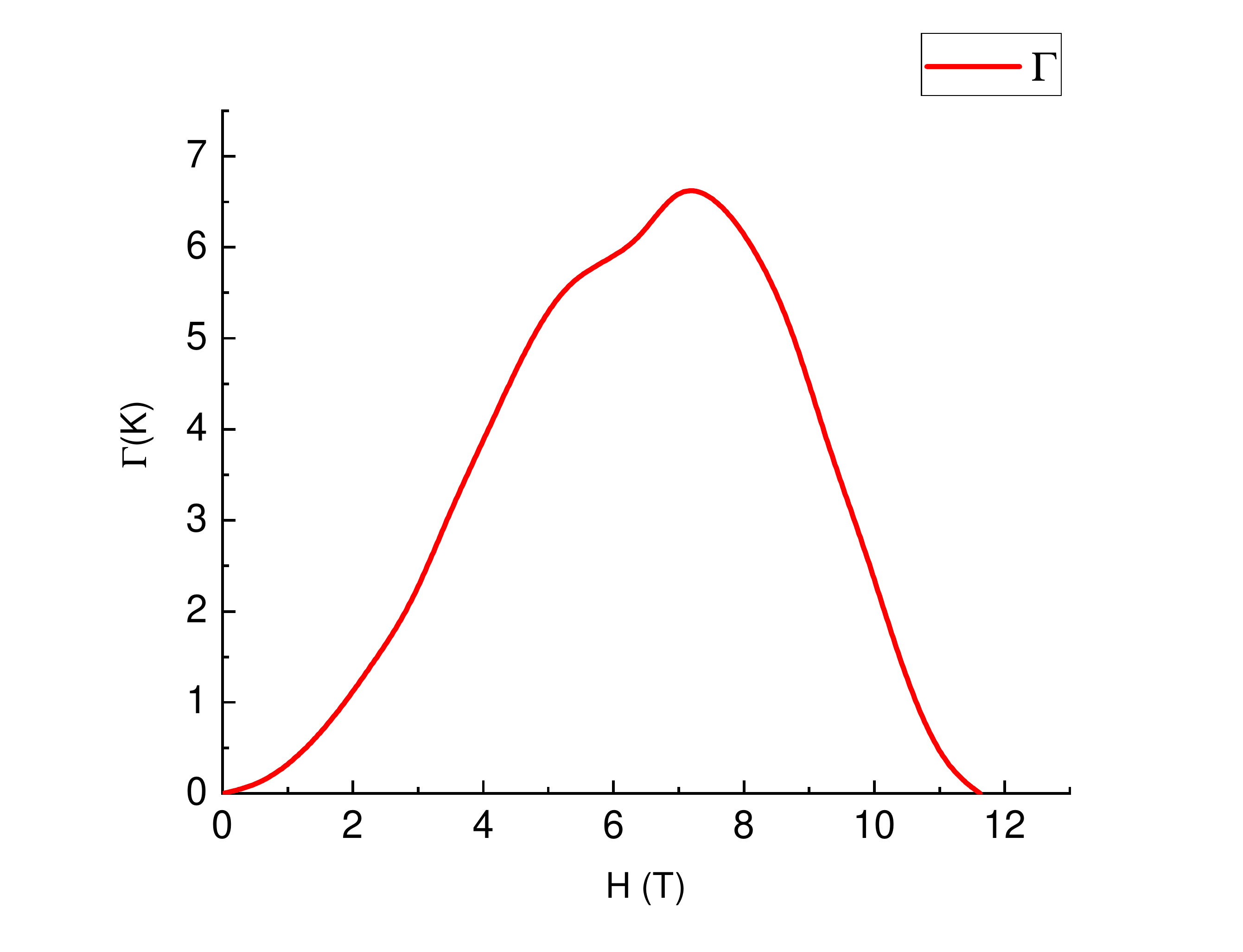}
  \caption{``Optical'' magnon with $\m{k}=\m{0}$ damping $\Gamma(h)$ as a function of external magnetic field. Parameters $D=8$ K, $J_c=2$ K, and  $J_a=0.2$ K were used.
\label{fig:Gw}}
\end{figure}

Our calculations show that $\Gamma(h)$ becomes zero at the critical fields $h_{c1}$, $h_{c2}$ and has a maximum with significant value in between (near the ordered phase center). This result is illustrated in Fig.~\ref{fig:Gw} for particular parameters set .

%


\subsection{Susceptibility}\label{spinsus}

The generalization of Eq.~\eqref{kubo} for the static transverse spin susceptibility is the sum of retarded Green's functions:
\begin{eqnarray}\label{susc1}
  \chi^{\perp}(\omega,\m{k})=-\dfrac{1}{2} \left( G^{R}_{\omega,\m{k}}+G^{R}_{-\omega,-\m{k}}+F^{R}_{\omega,\m{k}}+F^{R}_{-\omega,-\m{k}}\right), \nn \\
\end{eqnarray}
where one should take into account corrections from magnetic field $h_{x}$.

\begin{figure}[h]
  \centering
    \includegraphics[width=\linewidth]{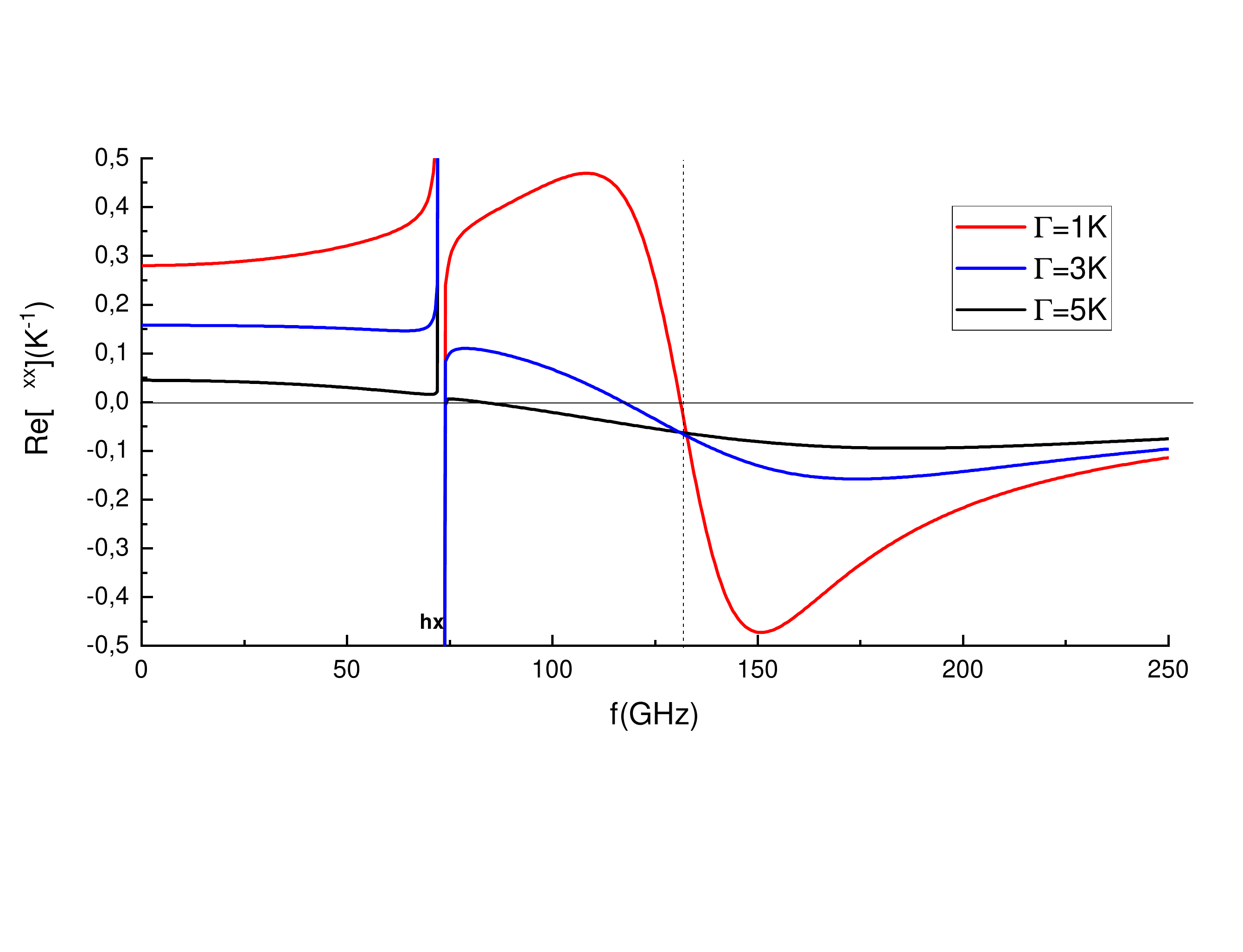}
  \caption{The real part of spin susceptibility $\chi^{\perp}$ at $H_{X}=7.3$T for different damping  constants $\Gamma$~\eqref{Gamma}. Parameters $\alpha = \pi/60$, $D=8$ K, $J_c=2$ K, and  $J_a=0.2$ K were used. 
\label{fig:Rechig}}
\end{figure}

In our calculations, in Eq.~\eqref{susc1} we use the first order in $1/S$ self-energies, which account for the umklapps. The result is the following (see Appendix~\ref{appendsusc} for the details):


\begin{eqnarray}\label{chixx}
&&\mathrm{Re} \left[ \chi^{\perp}(\omega, \m{k=0}) \right]=-\mathrm{Re} \Biggl\{ \frac{P(h)+2\overline{E}_{\mathbf{k}=0}-2\overline{B}_{\mathbf{k}=0}}{\omega^{2}-(\Delta-i\Gamma)^2}\nn\\  &&+\frac{K}{[(\omega+i\delta)^{2}-\Delta^{2}_{h_{x}}][\omega^{2}-(\Delta-i\Gamma)^2]} \Biggr\}.
\end{eqnarray}
Importantly, this formula has two types of poles, corresponding to ``stable'' acoustic  magnon  ($\delta \rightarrow +0$) and to damped optical magnon. So, an interplay between them determines uniform susceptibility as a function of frequency $\chi^\perp(\omega)$. Moreover, the magnitude of optical magnon damping $\Gamma$ plays a crucial role here, see Fig.~\ref{fig:Rechig}.

In the absence of the damping ($\Gamma=0$), the real part of the susceptibility at $\omega>0$ has two hyperbolic features associated with acoustic and optical magnon energies. However, when the damping grows up (what happens near the ordered phase center, see Fig.~\ref{fig:Gw}) the qualitative behavior of $\mathrm{Re} \, \chi^\perp(\omega)$ changes. Region of $\mathrm{Re} \, \chi^\perp(\omega) > 0$, pronounced near $\omega =\Delta$ at small $\Gamma$, fades away. Thus, in our analytical consideration, we observe the change of the susceptibility sign in that frequency domain, which was called ``dynamic diamagnetism'' in Ref.~\cite{smirnov2}. Moreover, our interpretation agrees with the one from this paper on a qualitative level, since two-magnon processes are the reason for large $\Gamma$ in our calculations.

\section{Discussion and Conclusions}
\label{SConc}

We study antiferromagnet with large single-ion easy-plane anisotropy in the tilted external field at small tilt angles in the ordered canted phase. In this case, the in-plane field component is small and we use linear response theory (Kubo formalism) to calculate the corresponding corrections to the ground state. This allows us to obtain corrections to the Hamiltonian, which appear due to the in-plane field component. Also, the latter breaks the rotational symmetry with respect to the tetragonal axis. Importantly, this leads to the gap in magnon spectrum emergence, which is non-monotonic under the external field variation. Theoretically, this feature is inherited from also nonmonotonic behavior of the energy of the optical magnon in the center of the Brillouin zone. It was studied in our previous paper~\cite{sherutesov} and connected with strong quantum fluctuations-induced renormalization of this energy.  Moreover, we show that our result for the gap field-dependence is in good qualitative agreement with the experimental findings of Ref.~\cite{smirnov}, where DTN was studied.

After establishing the ground state in the tilted field, our formalism also allows for studying transverse spin susceptibility. It is shown that there are two crucial points in this derivation. The first one is the  significant damping of the optical magnons, which is connected with decay on two magnons. The second one is accurate accounting for the umklapp terms in the Hamiltonian. 

As a result, using realistic DTN parameters, we theoretically explain the effect of dynamic diamagnetism, which was experimentally observed in Ref.~\cite{smirnov2}. Moreover, two-magnon processes plays important role in our derivation, which lies in agreement with the interpretation proposed in that paper.

It is pertinent to mention here, that the best fit of the experimental data was not our goal, and we use some realistic parameters of DTN just for illustration purposes. Furthermore, due to the quasi-1D nature of DTN, the dependence of the results on the parameter values is strong and the question of higher order corrections in the $1/S$ series can be also raised. Nevertheless, we find that our qualitative conclusions are true in a wide range of parameters. Moreover, we can report that the used parameters result in a semi-quantitative agreement with the experimental data.

Our theory and its results should be compared with the approach developed in Ref.~\cite{glazkov2020antiferromagnetic}. The advantages of the latter are its elegance and simple analytical formulas as the yield. However, it is valid only for magnetic fields close to the center of the ordered phase and relies on $J/D$ smallness. In our approach, the calculations are much more cumbersome, but the methods are more general. The latter allows us to address the transverse susceptibility in the same framework as the gap in the magnon spectrum.

\begin{acknowledgments}

We are grateful to A.I. Smirnov for stimulating discussions. The reported study was supported by the Foundation for the Advancement of Theoretical Physics and Mathematics ``BASIS''.

\end{acknowledgments}

\appendix

\section{Hamiltonian at $h_{x}=0$ \label{App1}}
\label{appendhx0}

Here we briefly remind the basic technique, which was used in our previous study~\cite{sherutesov} (see also Refs.~\cite{Hamer2010,sizanov2011antiferromagnet}). Hamiltonian~\eqref{hamiltonian} can be transformed into bosonic one using conventional Holstein-Primakoff spin-operators representation~\cite{holstein} in the approximate form (which allows taking into account first $1/S$ corrections):
\begin{eqnarray}\label{spinresp}
&S_{i}^{x^\prime}+iS_{i}^{y^\prime}=\sqrt{2S}a^{\dagger}_{i}\sqrt{1-\dfrac{a^{\dagger}_{i}a_{i}}{2S}}\approx\sqrt{2S}a^{\dagger}_{i}\left(1-\dfrac{a^{\dagger}_{i}a_{i}}{4S}\right),  \nonumber\\
&S_{i}^{x^\prime}-iS_{i}^{y^\prime}=\sqrt{2S}\sqrt{1-\dfrac{a^{\dagger}_{i}a_{i}}{2S}}a_{i}\approx\sqrt{2S}\left(1-\dfrac{a^{\dagger}_{i}a_{i}}{4S}\right)a_{i},  \nonumber\\
&S_{i}^{z^\prime}=-S+a^{\dagger}_{i}a_{i}
\end{eqnarray}
in a local coordinate frame, suitable for canted antiferromagnetic order description [$\m{k}_0= (\pi,\pi,\pi)$ is the AF vector]:
\begin{eqnarray}
\label{locspin}
& S_{i}^{x}=S_{i}^{x^\prime},\nonumber \\
&  S_{i}^{y}= S_{i}^{y^\prime}\cos{\theta}+S_{i}^{z^\prime}\exp(i\textbf{k}_{0}\textbf{R}_{i})\sin{\theta}, \nonumber\\
&  S_{i}^{z}= S_{i}^{z^\prime}\cos{\theta}-S_{i}^{y^\prime}\exp(i\textbf{k}_{0}\textbf{R}_{i})\sin{\theta}.\nonumber \\
\end{eqnarray}
As a result we have $\mathcal{H} = \sum^{4}_{n=0} \mathcal{H}_n$; each term contains $n$ Bose-operators. Explicitly:
\begin{eqnarray}
\mathcal{H}_{1}=i S^{3/2}\sqrt{N}\left(La_{\mathbf{k_{0}}}-La_{\mathbf{k_{0}}}^{\dagger}\right),\\
L=\dfrac{1}{\sqrt{2}}\left[(2J_{0}+2\widetilde{D})\cos\theta-\dfrac{h_{z}}{S} \right] \sin\theta.
\end{eqnarray}
This term can be eliminated by taking proper canting angle $\theta$, which depends on the external field. Next, the bilinear part is the following:
\begin{eqnarray}\label{h2}
&\mathcal{H}_{2}=S\sum_{\mathbf{k}}a_{\mathbf{k}}^{\dagger}a_{\mathbf{k}} E_{\mathbf{k}} +\frac{S}{2}\sum_{\mathbf{k}}(a_{\mathbf{k}}a_{-\mathbf{k}}+a_{\mathbf{k}}^{\dagger}a_{-\mathbf{k}}^{\dagger}) B_{\mathbf{k}}, \\
&E_{\mathbf{k}}=S(J_{0}+J_{\mathbf{k}})\cos^{2}\theta+S(J_{0}+D)\sin^{2}\theta,\\
&B_{\mathbf{k}}=S(J_{\mathbf{k}}-D)\sin^{2}\theta.
\end{eqnarray}
Here the Fourier transform of the exchange interaction reads
\be
  J_\m{k} = 2 \left[ J_c \cos{k_z} + J_a (\cos{k_x} + \cos{k_y})\right].
\ee
So, the magnon spectrum in the linear theory is given by:
\be \label{speclin}
\varepsilon_{\mathbf{k}}=\sqrt{{E^2_{\mathbf{k}}}-{B^2_{\mathbf{k}}}}.\label{linearspectrum}
\ee
Since the exchange interaction Fourier transform $J_{\m{k}}$ obeys the  property $J_{\m{0}} = - J_{\m{k}_0}$, this spectrum is gapless with $\varepsilon_{\m{k}_0}=0$ in agreement with the Goldstone's theorem. 

We also have the interaction part of the Hamiltonian. It consists of two terms, the first one is 
\begin{eqnarray}\label{h3}
\mathcal{H}_{3}&=&i\dfrac{\sqrt{S}\sin\theta\cos\theta}{4\sqrt{2N}} \sum_{123=k_{0}}(a_{1}^{\dagger}a_{2}^{\dagger}a_{-3}-a_{-3}^{\dagger}a_{2}a_{1})\dfrac{(V_\m{1}+V_\m{2})}{2},
\nonumber\\
\label{Vk}
V_{\mathbf{k}}&=&\left(2J_{\m{k}_0}-8J_{\mathbf{k}}+10D-\dfrac{h}{S\cos\theta}\right),
\end{eqnarray}
where we denote $\textbf{k}_{1}, \textbf{k}_{2},\textbf{k}_{3}$ as $1,2,3$, and the momentum  conservation law reads 
\be
\textbf{k}_{1}+\textbf{k}_{2}+\textbf{k}_{3}=\textbf{k}_{0}. \ee
The second one is the following:
\begin{eqnarray}
&&\mathcal{H}_{4}=\dfrac{1}{N}\sum_{1234=0}U_{\m{1234}}a_{1}^{\dagger}a_{2}^{\dagger}a_{-3}a_{-4}+\nonumber\\
&&+\dfrac{1}{N}\sum_{1234=0}V_{\m{123}}(a_{1}^{\dagger}a_{2}^{\dagger}a{3}^{\dagger}a_{-4}+a_{-4}^{\dagger}a_{3}a_{2}a_{1})
\ee
with the conventional momentum conservation law. Here
\be
U_{\m{1234}}&=&\Biggl[\dfrac{1-2\sin^{2}\theta}{2}J_{\m{4}-\m{1}}-\dfrac{\cos^{2}\theta}{4}(J_\m{1}+J_\m{4}) \\ && +D\left(1-\dfrac{3}{2}\sin^{2}\theta\right)\Bigg], \nn\\
V_{\m{123}}&=&\left[\dfrac{D\sin^{2}\theta}{4}-\dfrac{(J_\m{1}+J_\m{2}+J_\m{3})\sin^{2}\theta}{12}\right].
\end{eqnarray}

For non-linear corrections to magnon spectrum analysis, it is convenient to use normal ($G_{k}= \langle a_{\textbf{k}}, a_{\textbf{k}}^{\dagger} \rangle_{\omega}$) and anomalous ($F^\dagger_{k}=\langle a_{-\textbf{k}}^{\dagger}, a_{\textbf{k}}^{\dagger} \rangle_{\omega}$) Green's functions. The solution of the system of Dyson's equation for these functions reads
\begin{eqnarray} \label{Green}
&G_{k}=\dfrac{\omega+E_{\textbf{k}}+\Sigma_{-k}}{D_{\omega,k}},& \\ 
&F_{k}=-\dfrac{B_{\textbf{k}}+\Pi_{k}}{D_{\omega,k}},&\\
&D_{k}=\omega^2- \varepsilon^2_{\textbf{k}}-E_{\textbf{k}}(\Sigma_{k}+\Sigma_{-k})+2B_{\textbf{k}}\Pi_{k}+\nonumber\\
&+\omega(\Sigma_{k}-\Sigma_{-k}),&
\end{eqnarray}
where $\Sigma_{k}, \Pi_{k}$ are normal and anomalous self-energy parts, respectively. They can be calculated perturbatively.

\section{Corrections to the Hamiltonian at small  $h_{x}$.}
\label{appendhx}

When the external magnetic field is tilted, nonzero $h_x$ appears. It leads to uniform $\langle S^x \rangle$ component. In our theory, it is related to the parameter $\rho$. Nonzero $\rho$ provides various additional terms in the Hamiltonian, which can be altogether denoted as $\delta \mathcal{H}$.

Linear terms $\delta \mathcal{H}_1$ vanish when $\rho$ is properly chosen.

The bilinear part of $\delta \mathcal{H}_2$ can be divided onto normal and umklapp contributions [see Eqs.~\eqref{Hnorm} and~\eqref{Humk}]. Explicit expressions for the corresponding coefficients in $\delta \mathcal{H}^N_2$ are the following:
\begin{eqnarray}
  &\overline{E}_{\m{k}}=\dfrac{h_{x}\rho}{2S\sqrt{2NS}}+8U_{\m{0k0-k}}\dfrac{\rho^{2}}{NS}+12V_{\m{0k0}}\dfrac{\rho^{2}}{NS},\label{z}\\
  &\overline{B}_{\m{k}}=\dfrac{h_{x}\rho}{2S\sqrt{2NS}}+8U_{\m{kk00}}\dfrac{\rho^{2}}{NS}+12V_{\m{0k-k}}\dfrac{\rho^{2}}{NS}.\label{m}
\end{eqnarray}
These terms are analogous to the contributions in Eq.~\eqref{h2}. For $\delta \mathcal{H}^U_2$ the corresponding equations read
\begin{eqnarray}
&X_{\m{k}}=\sin\theta\cos\theta(V_{\m{0}}+V_{\m{k}})\dfrac{\rho}{4\sqrt{2NS}},\label{a}\\
&Y_{\m{k}}=\sin\theta\cos\theta(V_{\m{k}-\m{k}_{0}}+V_{\m{k}})\dfrac{\rho}{4\sqrt{2NS}}.\label{y}
\end{eqnarray}
Importantly, the umklapps provide corrections to the normal and anomalous self-energy parts $\sim \rho^2$ (they are combined in pairs, see Fig.~\ref{fig:Umklapp}):
\begin{eqnarray}
\label{sigmaU}\Sigma^{U}_{\m{k}}=X^{2}_{\m{k}}G_{\omega,\m{k-k_{0}}}+Y^{2}_{\m{k}}G_{-\omega,\m{k_{0}-k}}-\nonumber\\-X_{\m{k}}Y_{\m{k}}F_{\omega,\m{k-k_{0}}}-X_{\m{k}}Y_{\m{k}}F_{-\omega,\m{k_{0}-k}}\\
\nonumber\\
\label{piU}\Pi^{U}_{\m{k}}=X^{2}_{\m{k}}F_{\omega,\m{k-k_{0}}}+Y^{2}_{\m{k}}F_{-\omega,\m{k_{0}-k}}-\nonumber\\-X_{\m{k}}Y_{\m{k}}G_{\omega,\m{k-k_{0}}}-X_{\m{k}}Y_{\m{k}}G_{-\omega,\m{k_{0}-k}}
\end{eqnarray}
However, one can see that at the momentum $\m{k}=\m{k}_{0}$ condition $X_{\m{k}_{0}}=Y_{\m{k}_{0}}$ is satisfied, so $\Sigma^{U}_{\m{k_{0}}}=-\Pi^{U}_{\m{k_{0}}}$. This means that there is no correction to the energy of magnon with $\m{k}=\m{k}_{0}$ from the umklapp processes and only the normal terms are important in this context.

\section{Derivation of transverse susceptibility in tilted field}
\label{appendsusc}

Here, for calculations, it is also convenient to use normal and anomalous Green's functions:
\begin{equation}
  G_{k}= \langle b_{\textbf{k}}, b_{\textbf{k}}^{\dagger} \rangle_{\omega}; \, F^\dagger_{k}=\langle b_{-\textbf{k}}^{\dagger}, b_{\textbf{k}}^{\dagger} \rangle_{\omega},
\end{equation}
where $k=(\omega,\textbf{k})$. These functions obey the following system of Dyson's equations
\begin{eqnarray} \label{Dyson}
  G_{k}=G^{0}_{k}+G^{0}_{k} \, \Sigma_{k}\, G_{k}+G^{0}_{k}(B_{\textbf{k}}+\overline{B}_{\mathbf{k}}+\Pi_{k})F_{k}, \\ 
  F_{k}=G^{0}_{-k}(B_{\textbf{k}}+\overline{B}_{\mathbf{k}}+\Pi_{k})G_{k}+G^{0}_{-k} \, \Sigma_{-k} \, F_{k}, \nonumber
\end{eqnarray} 
where the bare Green's function reads
\be
  G^{0}=\dfrac{1}{\omega-E_{\textbf{k}}-\overline{E}_{\mathbf{k}}+i\delta}.
\end{eqnarray}
After solving the system of equations~\eqref{Dyson} we obtain the following expression for normal and anomalous Green's functions
\begin{eqnarray} \label{GreenE}
G_{k}=\dfrac{\omega+E_{\textbf{k}}+\overline{E}_{\mathbf{k}}+\Sigma_{k}}{D_{\omega,k}}; \\ F_{k}=-\dfrac{B_{\textbf{k}}+\overline{B}_{\mathbf{k}}+\Pi_{k}}{D_{\omega,k}}.
\end{eqnarray}
Here the denominator reads
\begin{eqnarray}
&D_{\omega,k}=\omega^2- \varepsilon^2_{\textbf{k}}-E_{\textbf{k}}(\Sigma_{k}+\Sigma_{-k}+2\overline{E}_{\mathbf{k}})\nonumber\\ & + 2B_{\textbf{k}}(\Pi_{k}+\overline{B}_{\mathbf{k}})+\omega(\Sigma_{k}-\Sigma_{-k}).\label{denominator}
\end{eqnarray}

To calculate the susceptibility~\eqref{susc1}, it is necessary to use the retarded Green's functions $G^{R}$ and $F^{R}$, which are related to the causal ones~\eqref{GreenE} by the following relation:
\begin{eqnarray}
G^{R}_{\omega, \mathbf{k}}=\theta(\omega)G_{\omega, \mathbf{k}}+\theta(-\omega)G^{*}_{\omega, \mathbf{k}},
\end{eqnarray}
and the same for $F$. The real parts of retarded functions are equal to the corresponding causal ones. So, we get
\begin{eqnarray}
\mathrm{Re} [\chi^{\perp}_{\omega, \m{k}}]= -\mathrm{Re} \Biggl\{ &&\frac{2E_{\mathbf{k}}+\Sigma_{k}+\Sigma_{-k}-2B_{\mathbf{k}}-2\Pi_{k}}{\omega^{2}-(\Delta-i\Gamma)^2}+\nn\\&&+\frac{2\overline{E}_{\mathbf{k}}-2\overline{B}_{\mathbf{k}}+2\Sigma^{U}_{\mathbf{k}}-2\Pi^{U}_{\mathbf{k}}}{\omega^{2}-(\Delta-i\Gamma)^2} \Biggr\}.
\end{eqnarray}
Using Eqs.~\eqref{sigmaU} and~\eqref{piU} at $\m{k}=0$ and Eq.~\eqref{Ph}, the real part of the uniform transverse spin susceptibility can be finally expressed as
\begin{eqnarray}
\mathrm{Re}[\chi^{\perp}_{\omega, \m{k=0}}]= -\mathrm{Re} \Biggl\{\frac{P(h)+2\overline{E}_{\mathbf{k}=0}-2\overline{B}_{\mathbf{k}=0}}{\omega^{2}-(\Delta-i\Gamma)^2}+\nn\\\frac{K}{[(\omega+i\delta)^{2}-\Delta^{2}_{h_{x}}][\omega^{2}-(\Delta-i\Gamma)^2]} \Biggr\},\nn\\
\end{eqnarray}
where 

\begin{eqnarray}
K=2(X_{\m{k=0}}+Y_{\m{k=0}})^{2}(\overline{E}_{\mathbf{k_{0}}}+\overline{B}_\mathbf{k_{0}}).
\end{eqnarray}

\bibliographystyle{abbrv}
\bibliography{References}

\begin{thebibliography}{10}

\bibitem{sachdev2011}
Subir Sachdev.
\newblock {\em Quantum Phase Transitions}.
\newblock Cambridge University Press, 2 edition, 2011.

\bibitem{Mila}
Frédéric Mila.
\newblock Quantum spin liquids.
\newblock {\em European Journal of Physics}, 21(6):499, 2000.

\bibitem{giamarchi2008bose}
Thierry Giamarchi, Christian R{\"u}egg, and Oleg Tchernyshyov.
\newblock Bose--einstein condensation in magnetic insulators.
\newblock {\em Nature Physics}, 4(3):198, 2008.

\bibitem{zheludev2013dirty}
Andrey Zheludev and Tommaso Roscilde.
\newblock Dirty-boson physics with magnetic insulators.
\newblock {\em Comptes Rendus Physique}, 14(8):740--756, 2013.

\bibitem{oosawa2002random}
Akira Oosawa and Hidekazu Tanaka.
\newblock Random bond effect in the quantum spin system (tl 1- x k x) cucl 3.
\newblock {\em Physical Review B}, 65(18):184437, 2002.

\bibitem{yu2012bose}
Rong Yu, Liang Yin, Neil~S Sullivan, JS~Xia, Chao Huan, Armando Paduan-Filho,
  Nei~F Oliveira~Jr, Stephan Haas, Alexander Steppke, Corneliu~F Miclea, et~al.
\newblock Bose glass and mott glass of quasiparticles in a doped quantum
  magnet.
\newblock {\em Nature}, 489(7416):379, 2012.

\bibitem{huvonen2012}
D.~H\"uvonen, S.~Zhao, M.~M\aa{}nsson, T.~Yankova, E.~Ressouche,
  C.~Niedermayer, M.~Laver, S.~N. Gvasaliya, and A.~Zheludev.
\newblock Field-induced criticality in a gapped quantum magnet with bond
  disorder.
\newblock {\em Phys. Rev. B}, 85:100410, Mar 2012.

\bibitem{Fisher}
Matthew~P.A. Fisher, Peter~B. Weichman, G.~Grinstein, and Daniel~S. Fisher.
\newblock Boson localization and the superfluid-insulator transition.
\newblock {\em Phys. Rev. B}, 40(1):546--570, 1989.

\bibitem{theorem}
L.~Pollet, N.~V. Prokof'ev, B.~V. Svistunov, and M.~Troyer.
\newblock Absence of a direct superfluid to mott insulator transition in
  disordered bose systems.
\newblock {\em Phys. Rev. Lett.}, 103:140402, 2009.

\bibitem{paduan2004}
A.~Paduan-Filho, X.~Gratens, and N.~F. Oliveira.
\newblock Field-induced magnetic ordering in
  ${\mathrm{nicl}}_{2}\ensuremath{\cdot}4\mathrm{SC}({\mathrm{nh}}_{2}{)}_{2}$.
\newblock {\em Phys. Rev. B}, 69:020405, Jan 2004.

\bibitem{Zvyagin2007}
S.~A. Zvyagin, J.~Wosnitza, C.~D. Batista, M.~Tsukamoto, N.~Kawashima,
  J.~Krzystek, V.~S. Zapf, M.~Jaime, N.~F. Oliveira, and A.~Paduan-Filho.
\newblock Magnetic excitations in the spin-1 anisotropic heisenberg
  antiferromagnetic chain system
  ${\mathrm{nicl}}_{2}\mathrm{\text{\ensuremath{-}}}4\mathrm{SC}({\mathrm{nh}}_{2}{)}_{2}$.
\newblock {\em Phys. Rev. Lett.}, 98:047205, Jan 2007.

\bibitem{sizanov2011antiferromagnet}
Alexey~V Sizanov and Arseny~V Syromyatnikov.
\newblock Antiferromagnet with two coupled antiferromagnetic sublattices in a
  magnetic field.
\newblock {\em Journal of Physics: Condensed Matter}, 23(14):146002, 2011.

\bibitem{Sizanov2011}
A.~V. Sizanov and A.~V. Syromyatnikov.
\newblock Bosonic representation of quantum magnets with large single-ion
  easy-plane anisotropy.
\newblock {\em Phys. Rev. B}, 84:054445, Aug 2011.

\bibitem{Povarov2017}
K.~Yu. Povarov, A.~Mannig, G.~Perren, J.~S. M\"oller, E.~Wulf, J.~Ollivier, and
  A.~Zheludev.
\newblock Quantum criticality in a three-dimensional spin system at zero field
  and pressure.
\newblock {\em Phys. Rev. B}, 96:140414, Oct 2017.

\bibitem{Orlova_2018}
A.~Orlova, H.~Mayaffre, S.~Kr\"amer, M.~Dupont, S.~Capponi, N.~Laflorencie,
  A.~Paduan-Filho, and M.~Horvati\ifmmode~\acute{c}\else \'{c}\fi{}.
\newblock Detection of a disorder-induced bose-einstein condensate in a quantum
  spin material at high magnetic fields.
\newblock {\em Phys. Rev. Lett.}, 121:177202, Oct 2018.

\bibitem{zapf2006}
V.~S. Zapf, D.~Zocco, B.~R. Hansen, M.~Jaime, N.~Harrison, C.~D. Batista,
  M.~Kenzelmann, C.~Niedermayer, A.~Lacerda, and A.~Paduan-Filho.
\newblock Bose-einstein condensation of $s=1$ nickel spin degrees of freedom in
  ${\mathrm{nicl}}_{2}\mathrm{\text{\ensuremath{-}}}4\mathrm{SC}({\mathrm{nh}}_{2}{)}_{2}$.
\newblock {\em Phys. Rev. Lett.}, 96:077204, Feb 2006.

\bibitem{batyev1984}
EG~Batyev and LS~Braginsky.
\newblock Antiferromagnet in a strong magnetic-field-analogy with a bose-gas.
\newblock {\em Sov. Phys. JETP}, 60(4):781, 1984.

\bibitem{batyev1985}
EG~Batyev.
\newblock Antiferromagnet of arbitrary spin in a strong magnetic field.
\newblock {\em Sov. Phys. JETP}, 62(1):173, 1985.

\bibitem{yin2008}
L.~Yin, J.~S. Xia, V.~S. Zapf, N.~S. Sullivan, and A.~Paduan-Filho.
\newblock Direct measurement of the bose-einstein condensation universality
  class in
  ${\mathrm{nicl}}_{2}\mathrm{\text{\ensuremath{-}}}4\mathrm{SC}({\mathrm{nh}}_{2}{)}_{2}$
  at ultralow temperatures.
\newblock {\em Phys. Rev. Lett.}, 101:187205, Oct 2008.

\bibitem{smirnov}
S.~A. Zvyagin, J.~Wosnitza, A.~K. Kolezhuk, V.~S. Zapf, M.~Jaime,
  A.~Paduan-Filho, V.~N. Glazkov, S.~S. Sosin, and A.~I. Smirnov.
\newblock Spin dynamics of
  $\mathrm{Ni}{\mathrm{cl}}_{2}\text{\ensuremath{-}}4\mathrm{S}\mathrm{C}{(\mathrm{N}{\mathrm{H}}_{2})}_{2}$
  in the field-induced ordered phase.
\newblock {\em Phys. Rev. B}, 77:092413, Mar 2008.

\bibitem{smirnov2}
T.~A. Soldatov, A.~I. Smirnov, K.~Yu. Povarov, A.~Paduan-Filho, and
  A.~Zheludev.
\newblock Microwave dynamics of the stoichiometric and bond-disordered
  anisotropic $s=1$ chain antiferromagnet
  ${\mathrm{nicl}}_{2}\text{\ensuremath{-}}4\mathrm{SC}{({\mathrm{NH}}_{2})}_{2}$.
\newblock {\em Phys. Rev. B}, 101:104410, Mar 2020.

\bibitem{sherutesov}
Artemiy~S Sherbakov and Oleg~I Utesov.
\newblock Magnon spectrum and electron spin resonance in antiferromagnet with
  large single-ion easy plane anisotropy.
\newblock {\em Journal of Magnetism and Magnetic Materials}, 518:167390, 2021.

\bibitem{Lopez1963}
Amparo Lopez-Castro and Mary~R. Truter.
\newblock 245. the crystal and molecular structure of
  dichlorotetrakisthioureanickel{,} [(nh2)2cs]4nicl2.
\newblock {\em J. Chem. Soc.}, pages 1309--1317, 1963.

\bibitem{holstein}
T.~Holstein and H.~Primakoff.
\newblock Field dependence of the intrinsic domain magnetization of a
  ferromagnet.
\newblock {\em Phys. Rev.}, 58:1098--1113, Dec 1940.

\bibitem{Hamer2010}
C.~J. Hamer, O.~Rojas, and J.~Oitmaa.
\newblock Spin-wave analysis of the spin-1 heisenberg antiferromagnet with
  uniaxial single-ion anisotropy in a field.
\newblock {\em Phys. Rev. B}, 81:214424, Jun 2010.

\bibitem{glazkov2020antiferromagnetic}
Vasily~Nikolaevich Glazkov.
\newblock Antiferromagnetic resonance in a spin-gap magnet with strong
  single-ion anisotropy.
\newblock {\em JETP Letters}, 112(10):647--650, 2020.

\end{thebibliography}

\end{document}